\documentclass[ALICE,manyauthors]{cernphprep}
\usepackage[comma,square,numbers,sort&compress]{natbib}
\usepackage{hyperref}
\usepackage{lineno}
\usepackage{xspace}
\usepackage[utf8]{inputenc}
\usepackage{siunitx}
\usepackage{verbatim}
\usepackage[T1]{fontenc}
\usepackage{orcidlink}

\begin{document}
%

\newcommand{\pp}           {pp\xspace}
\newcommand{\ppbar}        {\mbox{$\mathrm {p\overline{p}}$}\xspace}
\newcommand{\XeXe}         {\mbox{Xe--Xe}\xspace}
\newcommand{\PbPb}         {\mbox{Pb--Pb}\xspace}
\newcommand{\pPb}          {\mbox{p--Pb}\xspace}
\newcommand{\AuAu}         {\mbox{Au--Au}\xspace}
\newcommand{\dAu}          {\mbox{d--Au}\xspace}

\newcommand{\snn}          {\ensuremath{\sqrt{s_{\mathrm{NN}}}}\xspace}
\newcommand{\pt}           {\ensuremath{p_{\rm T}}\xspace}
\newcommand{\meanpt}       {$\langle p_{\mathrm{T}}\rangle$\xspace}
\newcommand{\ycms}         {\ensuremath{y_{\rm CMS}}\xspace}
\newcommand{\ylab}         {\ensuremath{y_{\rm lab}}\xspace}
\newcommand{\etarange}[1]  {\mbox{$\left | \eta \right |~<~#1$}}
\newcommand{\yrange}[1]    {\mbox{$\left | y \right |~<~#1$}}
\newcommand{\dndy}         {\ensuremath{\mathrm{d}N_\mathrm{ch}/\mathrm{d}y}\xspace}
\newcommand{\dndeta}       {\ensuremath{\mathrm{d}N_\mathrm{ch}/\mathrm{d}\eta}\xspace}
\newcommand{\avdndeta}     {\ensuremath{\langle\dndeta\rangle}\xspace}
\newcommand{\dNdy}         {\ensuremath{\mathrm{d}N_\mathrm{ch}/\mathrm{d}y}\xspace}
\newcommand{\Npart}        {\ensuremath{N_\mathrm{part}}\xspace}
\newcommand{\Ncoll}        {\ensuremath{N_\mathrm{coll}}\xspace}
\newcommand{\dEdx}         {\ensuremath{\textrm{d}E/\textrm{d}x}\xspace}
\newcommand{\RpPb}         {\ensuremath{R_{\rm pPb}}\xspace}

\newcommand{\nineH}        {$\sqrt{s}~=~0.9$~Te\kern-.1emV\xspace}
\newcommand{\seven}        {$\sqrt{s}~=~7$~Te\kern-.1emV\xspace}
\newcommand{\twoH}         {$\sqrt{s}~=~0.2$~Te\kern-.1emV\xspace}
\newcommand{\twosevensix}  {$\sqrt{s}~=~2.76$~Te\kern-.1emV\xspace}
\newcommand{\five}         {$\sqrt{s}~=~5.02$~Te\kern-.1emV\xspace}
\newcommand{\twosevensixnn}{$\sqrt{s_{\mathrm{NN}}}~=~2.76$~Te\kern-.1emV\xspace}
\newcommand{\fivenn}       {$\sqrt{s_{\mathrm{NN}}}~=~5.02$~Te\kern-.1emV\xspace}
\newcommand{\onethirtynn}       {$\sqrt{s_{\mathrm{NN}}}~=~130$~Ge\kern-.1emV\xspace}
\newcommand{\twohundrednn}       {$\sqrt{s_{\mathrm{NN}}}~=~200$~Ge\kern-.1emV\xspace}
\newcommand{\seventev}        {$\sqrt{s}~=~7$~Te\kern-.1emV\xspace}

\newcommand{\LT}           {L{\'e}vy-Tsallis\xspace}
\newcommand{\GeVc}         {\ensuremath{\mathrm{GeV}/c}\xspace}
\newcommand{\MeVc}         {Me\kern-.1emV/$c$\xspace}
\newcommand{\GeVmass}      {Ge\kern-.2emV/$c^2$\xspace}
\newcommand{\MeVmass}      {Me\kern-.2emV/$c^2$\xspace}
\newcommand{\lumi}         {\ensuremath{\mathcal{L}}\xspace}
\newcommand{\geantf}       {\ensuremath{\textsc{Geant4}}\xspace}

\newcommand{\ITS}          {\rm{ITS}\xspace}
\newcommand{\TOF}          {\rm{TOF}\xspace}
\newcommand{\ZDC}          {\rm{ZDC}\xspace}
\newcommand{\ZDCs}         {\rm{ZDCs}\xspace}
\newcommand{\ZNA}          {\rm{ZNA}\xspace}
\newcommand{\ZNC}          {\rm{ZNC}\xspace}
\newcommand{\SPD}          {\rm{SPD}\xspace}
\newcommand{\SDD}          {\rm{SDD}\xspace}
\newcommand{\SSD}          {\rm{SSD}\xspace}
\newcommand{\TPC}          {\rm{TPC}\xspace}
\newcommand{\TRD}          {\rm{TRD}\xspace}
\newcommand{\VZERO}        {\rm{V0}\xspace}
\newcommand{\VZEROA}       {\rm{V0A}\xspace}
\newcommand{\VZEROC}       {\rm{V0C}\xspace}
\newcommand{\Vdecay} 	   {\ensuremath{V^{0}}\xspace}

\newcommand{\ee}           {\ensuremath{e^{+}e^{-}}} 
\newcommand{\pip}          {\ensuremath{\pi^{+}}\xspace}
\newcommand{\pim}          {\ensuremath{\pi^{-}}\xspace}
\newcommand{\kap}          {\ensuremath{\rm{K}^{+}}\xspace}
\newcommand{\kam}          {\ensuremath{\rm{K}^{-}}\xspace}
\newcommand{\pbar}         {\ensuremath{\rm\overline{p}}\xspace}
\newcommand{\kzero}        {\ensuremath{{\rm K}^{0}_{\rm{S}}}\xspace}
\newcommand{\lmb}          {\ensuremath{\Lambda}\xspace}
\newcommand{\almb}         {\ensuremath{\overline{\Lambda}}\xspace}
\newcommand{\Om}           {\ensuremath{\Omega^-}\xspace}
\newcommand{\Mo}           {\ensuremath{\overline{\Omega}^+}\xspace}
\newcommand{\X}            {\ensuremath{\Xi^-}\xspace}
\newcommand{\Ix}           {\ensuremath{\overline{\Xi}^+}\xspace}
\newcommand{\Xis}          {\ensuremath{\Xi^{\pm}}\xspace}
\newcommand{\Oms}          {\ensuremath{\Omega^{\pm}}\xspace}

\newcommand{\SigmaInelAntiD}{\ensuremath{\sigma_{\rm inel}(\overline{\rm d})}\xspace}
\newcommand{\SigmaInelAntiP}{\ensuremath{\sigma_{\rm inel}(\overline{\rm p})}\xspace}
\newcommand{\SigmaInelD}{\ensuremath{\sigma_{\rm inel}(\rm d)}\xspace}
\newcommand{\SigmaInelAntiT}{\ensuremath{\sigma_{\rm inel}(^{3}\overline{\rm H})}\xspace}
\newcommand{\SigmaInelP}{\ensuremath{\sigma_{\rm inel}(\rm p)}\xspace}
\newcommand{\RatioAntid}{\ensuremath{\overline{\rm d}/{\rm d}}\xspace}
\newcommand{\RatioAntit}{\ensuremath{\overline{{^3\rm H}}/{\rm {^3H}}}\xspace}
\newcommand{\RatioAntip}{\ensuremath{\overline{\rm p}/{\rm p}}\xspace}
\newcommand{\Antip}{\ensuremath{\overline{\rm p}}\xspace}
\newcommand{\Antid}{\ensuremath{\overline{\rm d}}\xspace}
\newcommand{\AntiHeThree}{\ensuremath{^{3}\overline{\rm He}}\xspace}
\newcommand{\HeThree}{\ensuremath{^{3}{\rm He}}\xspace}
\newcommand{\AntiHeFour}{\ensuremath{^{4}\overline{\rm He}}\xspace}
\newcommand{\HeFour}{\ensuremath{^{4}{\rm He}}\xspace}
\newcommand{\AntiHThree}{\ensuremath{^{3}\overline{\rm H}}\xspace}
\newcommand{\antit}{\ensuremath{^{3}\overline{\rm H}}\xspace}

\newcommand{\eg}{e.\,g.\xspace}
\newcommand{\ie}{i.\,e.\xspace}

\newcommand{\mb}{\ensuremath{~\si{\milli\barn}}\xspace}
\newcommand{\mub}{\ensuremath{~\si{\micro\barn}}\xspace}

\newcommand{\mbinv} {\ensuremath{~\si{\per\milli\barn}}\xspace}
\newcommand{\mubinv} {\ensuremath{~\si{\per\micro\barn}}\xspace}
\newcommand{\nbinv} {\ensuremath{~\si{\per\nano\barn}}\xspace}
\newcommand{\pbinv} {\ensuremath{~\si{\per\pico\barn}}\xspace}
\newcommand{\fbinv} {\ensuremath{~\si{\per\femto\barn}}\xspace}
\newcommand{\Lumi}{\ensuremath{\mathcal{L}}\xspace}
\newcommand{\pvert} {\ensuremath{p_{\mathrm{primary}}}\xspace}
\newcommand{\ahe} {\ensuremath{^3\mathrm{\overline{He}}}\xspace}

\newcommand{\detailtexcount}[1]{%
  \immediate\write18{texcount -merge -sum -q #1.tex output.bbl > #1.wcdetail }%
  \verbatiminput{#1.wcdetail}%
  \newpage
}

\newcommand{\quickwordcount}[1]{%
  \immediate\write18{texcount -1 -sum -merge -q #1.tex output.bbl > #1-words.sum }%
  \input{#1-words.sum} words%
}

\newcommand{\quickcharcount}[1]{%
  \immediate\write18{texcount -1 -sum -merge -char -q #1.tex output.bbl > #1-chars.sum }%
  \input{#1-chars.sum} characters (not including spaces)%
}

\begin{titlepage}
\PHyear{2023}       
\PHnumber{118}      
\PHdate{22 June}  

\title{Measurement of the low-energy antitriton inelastic cross section}   

\Collaboration{ALICE Collaboration\thanks{See Appendix~\ref{app:collab} for the list of collaboration members}}
\ShortAuthor{ALICE Collaboration} 

\begin{abstract}

In this Letter, the first measurement of the inelastic cross section for antitriton--nucleus interactions is reported, covering the momentum range of $0.8 \leq p < 2.4$ \GeVc.
The measurement is carried out using data recorded with the ALICE detector in pp and Pb--Pb collisions at a centre-of-mass energy per nucleon of 13 TeV and 5.02 TeV, respectively. The detector material serves as an absorber for antitriton nuclei.
The raw yield of (anti)triton nuclei measured with the ALICE apparatus is compared to the results from detailed ALICE simulations based on the \geantf toolkit for the propagation of \mbox{(anti)}particles through matter, allowing one to quantify the inelastic interaction probability in the detector material. This analysis complements the measurement of the inelastic cross section of antinuclei up to $A=3$ carried out by the ALICE Collaboration, and demonstrates the feasibility of the study of the isospin dependence of inelastic interaction cross section with the analysis techniques presented in this Letter.

\end{abstract}
\end{titlepage}

\setcounter{page}{2} 


\section{Introduction}
\label{Sect:Introduction}

The properties of light antinuclei (such as antideuteron $\overline{\rm d}$, antihelium $^{3}\overline{\rm He}$ and antitriton $^{3}\overline{\rm H}$) have been a subject of various studies at accelerators in the last few decades.
These objects, composed of antiprotons and antineutrons, are stable in vacuum, but annihilate when coming into contact with normal matter. So far, light antinuclei have only been observed in high-energy particle collisions at various fixed-target and collider experiments, from the AGS~\cite{Bennett:1998be, Ahle:1999in, Armstrong:2000gz, Armstrong:2000gd}, to the SPS~\cite{ Ambrosini:1997bf}, RHIC~\cite{Adler:2001prl,Adler:2004uy, Arsene:2010px,Agakishiev:2011ib,Adamczyk:2016gfs,Adam:2019wnb}, and the LHC~\cite{Adam:2015pna,nuclei_pp_PbPb,dbar_coalescence_ALICE,deuteron_PbPb_276TeV,nuclei_pp,Acharya:2017bso,deuteron_pPbALICE,3He_pPb,deuteron_pp_13TeV,deuteron_flow_5TeV,helium3_flow_5TeV,deuteron_jets,nuclei_pp_5TeV,hypertriton_pPb_ALICE,nuclei_pp_13TeV,ALICE:2022weh,ALICE:2022boh,ALICE:2022ugx}. Their production mechanism has been actively studied in the context of matter--antimatter asymmetry 
and of understanding the formation of composite nuclear objects~\cite{Andronic:2010qu,Cleymans:2011pe,Vovchenko:2018fiy,Butler:1963pp,Scheibl:1998tk,nuclei_pp_5TeV}. However, until very recently, the studies of their inelastic interactions with matter have been difficult, since obtaining an isolated beam of light antinuclei with precisely determined momentum comprises formidable challenges. In fact, only the antideuteron inelastic cross section had been measured at total momenta of 13.3 GeV/$c$~\cite{Denisov:1971im} and 25 GeV/$c$~\cite{Binon:1970yu}.

Recently, new measurements of the inelastic cross sections of antinuclei with matter at low momenta ($p\leq 8$ \GeVc) were reported by the ALICE Collaboration, both for $\overline{\rm d}$~\cite{antideuteronXS} and \ahe~\cite{antiHe3XS}. While these measurements were mainly motivated by their impact on astrophysical dark-matter searches, they also allow a more detailed look into the physics of low-energy antinucleus--nucleus interactions, which can be used to improve Monte Carlo (MC) simulations of particle physics detectors and to better understand the properties of these composite objects at low kinetic energies. More accurate MC simulations of inelastic interactions with the detector material also allow one to improve the measurements of the production yields of antinuclei~\cite{ALICE:2022amd,nuclei_pp_5TeV,nuclei_pp_13TeV}. However, one aspect which so far has not been accessible with these measurements is the isospin dependence of the inelastic cross section of composite antinuclei. This effect is difficult to evaluate since the lightest isospin-partner antinuclei are \ahe\ and \AntiHThree, which are only produced in limited quantities in high-energy hadronic collisions at accelerators~\cite{nuclei_pp_13TeV}.

In this Letter the first measurement of the antitriton--nucleus inelastic interaction cross section \SigmaInelAntiT is reported, which complements the measurements of the inelastic cross sections for light antinuclei up to $A=3$~\cite{antideuteronXS,antiHe3XS}. The results are compared to the calculations based on the Glauber approach as implemented in the \geantf\ toolkit, which is widely used in particle physics for the propagation of particles through the detector material~\cite{Agostinelli:2002hh,Allison:2006ve}. The \SigmaInelAntiT results are also compared with the measurement of the \ahe\ inelastic cross section and demonstrate the feasibility to investigate the isospin dependence of inelastic interactions of antinuclei.

\section{Experimental apparatus and data sample}
\label{Sect:experiment}

A detailed description of the performance of the ALICE detector and its subsystems can be found in Refs.~\cite{Aamodt:2008zz,Abelev:2014ffa}. The detectors used in this analysis for tracking and particle identification are briefly described in the following paragraph.

Trajectories of charged particles are reconstructed in the ALICE central barrel with the Inner Tracking System (ITS)~\cite{ALICE:1999cls} and the Time Projection Chamber (TPC)~\cite{Alme:2010ke}. Both these detectors cover the full azimuthal angle within the pseudorapidity interval of $|\eta|<$ 0.9 and reside within a solenoid which provides a homogeneous magnetic field of $\SI{0.5}{\tesla}$ along the beam axis. The ITS comprises six cylindrical layers of silicon detectors which are located at radial distances from the beam axis between $\SI{3.9}{\centi\meter}$ and $\SI{43}{\centi\meter}$ and provides important information to reconstruct the primary collision vertex and to aid in charged particle tracking. It is surrounded by the TPC -- the main tracking device in the ALICE central barrel which is a 5 m long cylindrical gaseous detector extending from 85 cm to 247 cm in the radial direction. The TPC provides up to 159 spatial points for the reconstruction of charged-particle tracks and for particle identification (PID) through the measurement of the specific energy loss $\mathrm{d} E/\mathrm{d} x$ in the gas volume. The PID is complemented by the Time-of-Flight (TOF) detector~\cite{Akindinov:2013tea} located at the radial distance of 3.7 m from the beam axis. The TOF measures the arrival time of particles relative to the event collision time which is provided either by the TOF itself or by the T0 detectors, two arrays of Cherenkov counters located at forward rapidities~\cite{Adam:2016ilk}. Together with the momentum of the particle obtained from the track curvature, the time-of-flight measurement allows one to determine the particle's mass. Finally, a stainless steel space-frame supporting structure and the Transition Radiation Detector (TRD) are located between the TPC and the TOF; while neither the space-frame nor the TRD is used directly in this analysis, they significantly contribute to the total material budget in which antinuclei can interact inelastically~\cite{antiHe3XS,antideuteronXS}.

The results presented in this Letter are based on the data collected during the 2016, 2017, and 2018 LHC operation with proton beams at a centre-of-mass energy $\sqrt{s}$ = 13 TeV with the high-multiplicity (HM) trigger, and the 2018 Pb--Pb campaign at \snn = 5.02 TeV with the minimum bias (MB), central (0--10\%) or semi-central (30--50\%) event triggers. The central event trigger corresponds to the selection of 10\% of all inelastic events with the highest signal amplitude in the V0 detectors (0--10\% centrality~\cite{ALICE:2013hur}), whereas semi-central trigger is tuned to select events within 30--50\% centrality range. Collision events are selected by using the information from the V0 detectors~\cite{VZEROPerformance}, which consist of two plastic scintillator arrays located on both sides of the interaction point at forward and backward pseudorapidities. The V0 detectors are also used to reject other sources of tracks such as beam--gas interactions and interactions within the beampipe. For the MB event trigger, coincident signals in both V0 scintillators are required to be synchronous with the beam crossing time defined by the LHC clock. In order to trigger on events with high charged-particle multiplicities, the total signal amplitude measured in the V0 detector is additionally required to exceed a given threshold for a given event activity. This selects the 0.17\% of events with the highest multiplicity in the V0 detectors in pp collisions. An additional cut is made on the z position of the primary vertex ($|Vtx_z|<10$ cm).

In total, about 10$^9$ events from the pp data sample are selected for further analysis, while for the Pb--Pb dataset about $230 \times 10^6$ events are analysed.

\section{Data analysis}
\label{Sect:dataanalysis}
Once produced in the initial collisions between protons or heavy ions, antitriton nuclei traverse the detector, with some of them interacting inelastically with the detector material. In a traditional fixed-target experiment, the measurement of a cross section implies the availability of a beam of the particle of interest, which traverses the target in order to determine the loss of particles.  Due to the unfeasibility of isolating a beam of low energy antinuclei, the measurements presented in this work rely on ratios that are sensitive to this loss without depending on the absolute number of produced antinuclei. Two methods have been applied to evaluate the inelastic cross section of antitritons, \SigmaInelAntiT, similarly to what has been done in Refs.~\cite{antiHe3XS, antideuteronXS} for the \ahe\ and $\overline{\mathrm{d}}$ inelastic cross section measurements.

The first method, namely the antibaryon-to-baryon method, was used to analyse the high-multiplicity pp collisions and is based on the comparison of measured \AntiHThree and $^3$H yields. Due to the extra inelastic processes which occur as particles traverse the detector, less \AntiHThree is detected than its matter counterpart. This loss is quantified by comparison with detailed MC simulations of the ALICE detector. This method makes use of the fact that the relative amounts of matter and antimatter produced at LHC energies are almost equal~\cite{primordial_ratio1, primordial_ratio2}, and for  $^3\overline{\mathrm{H}}$ and $^3\mathrm{H}$ can be calculated from the antiproton-to-proton ratio~\cite{primordial_ratio2, Aamodt:2010dx} to be 0.994 $_{-0.045}^{+0.006}$ . Thus, the antiparticle-to-particle ratio is sensitive to the loss of antinuclei. This method allows one to extend the measurement of \AntiHThree up to $\pvert = 2.4$ GeV/$c$, but is unfeasible for application in Pb--Pb collisions due to copious $^3$H background from spallation processes~\cite{3He_pPb}.

The second method, applied in the analysis of Pb--Pb data, relies on the measurement of the amount of raw \AntiHThree reconstructed in the TOF detector and of those reconstructed in the TPC. This is the TOF-to-TPC method employed in Ref.~\cite{antiHe3XS}. In this case, the number of \AntiHThree nuclei tagged in the TPC and measured in the TOF is compared to that of \AntiHThree reconstructed in the TPC, which allows one to quantify the inelastic processes occurring in the material between the TPC and the TOF detectors. This is an almost direct analogy of a fixed target experiment, as the particles identified in the TPC act as the beam, while the ones reaching the TOF represent the particles surviving after the target. In order to extract the inelastic cross section while avoiding any bias due to finite detector acceptance and tracking efficiencies, data are compared to MC simulations that reproduce the conditions of the data taking. This method is applicable only in the momentum range in which \AntiHThree can be clearly identified in both TPC and TOF detectors ($0.9 < \pvert < 1.5$ GeV/$c$) and is used in the analysis of Pb--Pb data due to the larger yield of produced \AntiHThree nuclei compared to pp collisions.
\\ 
For both methods, the experimental results are compared with the corresponding ratios evaluated by means of a full-scale MC \geantf simulation with varied \SigmaInelAntiT cross sections, as described in section~\ref{Sect:MC}.
In both cases, using ratios instead of individual particle yields allows one to extract the antitriton inelastic cross section independently from its production cross section. Further details on the data analysis are described in the following sections.

\subsection{Track selection and particle identification}
\label{Sect:pid}
(Anti)$^{3}$H candidates are selected from a sample of charged-particle tracks reconstructed in the ITS and TPC in the pseudorapidity range $|\eta| < 0.8 \ (0.75)$ for the Pb--Pb (pp) data sample, and at midrapidity with |$y$| < 0.5 for both samples. Several track quality criteria are applied, such as a minimum number of clusters in the TPC of at least 100 out of a maximum of 159, and at least 2 in the ITS, with at least one cluster located in any of the two innermost ITS layers. Furthermore, the number of TPC crossed rows is constrained to be more than 70.
A good quality of the track fit is achieved by requiring the $\chi^{2}$ per TPC reconstructed point to be less than 2.5 (4) for the Pb--Pb (pp) data sample. These cuts are stricter in the Pb--Pb data sample due to the higher track density environment.
The number of TPC clusters used in the calculation of the specific energy loss (\dEdx) is required to be larger than 100 (55) to ensure a good \dEdx resolution in the Pb--Pb (pp) dataset.
The contribution from secondary tracks in the pp data sample is reduced by selecting a maximum distance of closest approach (DCA) to the primary vertex in the transverse plane ($\mathrm{DCA}_{xy}$) and in the longitudinal direction ($\mathrm{DCA}_{z}$) lower than 0.1 cm. 

The (anti)triton candidate tracks are identified using the information on the specific energy loss \dEdx in the TPC gas volume, which is complemented by the information on the time-of-flight measurement in the TOF detector.
The $n\sigma^{\rm TPC}$ variable represents the PID response in the TPC expressed in terms of the deviation between the measured and expected \dEdx, normalised by the detector resolution $\sigma$. The expected \dEdx is computed with a parameterised Bethe--Bloch formula~\cite{Abelev:2014ffa}.
For the antibaryon-to-baryon method, \mbox{(anti)}tritons are selected in the TPC by applying the selection criterion $|n\sigma^{\rm TPC}| < 3.0$, once the requirement on the squared mass hypothesis of the (anti)tritons $|\mathrm{m}_{\mathrm{TOF}}^2 - \mathrm{m}_{^3\mathrm{\overline{H}}}^2|<2$ (GeV/$c^2$)$^2$  measured with the TOF detector is fulfilled. Candidate tracks which do not reach the TOF are not considered in the analysis.
This selection is sufficient to obtain a purity close to 100\% for \mbox{(anti)}tritons in the full momentum range explored in this analysis (1.3 $<p<$ 2.4 \GeVc).

For the TOF-to-TPC method, \AntiHThree candidates need to be reconstructed in both TPC and TOF detectors in the same momentum intervals, hence the analysis is restricted to the momentum range of 0.9 $<p<$ 1.5 \GeVc. Antitriton candidates are selected in the TPC by applying the selection of $-2.0<n\sigma^{\rm TPC}< 3.5$, and then matched to TOF hits. The asymmetric selection interval allows one to suppress the contamination due to other particle species, misidentified as \AntiHThree in the TPC, with lower \dEdx signal. Any residual contamination in the $n\sigma^{\rm TPC}$ distributions is fitted with an exponential function and subtracted. Such contribution is negligible at low momentum and amounts to < 1\% in the momentum interval 1.3 $<p<$ 1.5 \GeVc.
Similarly to the $n\sigma^{\rm TPC}$, the $n\sigma^{\rm TOF}$ variable represents the deviation between the measured and expected time-of-flight of antitritons, normalised by the resolution of the time-of-flight measurement. 
In the TOF, \AntiHThree nuclei are selected in the range of $-3.0<n\sigma^{\rm TOF}< 4.0$. 
The reason for an asymmetric interval is the presence of an exponential tail towards higher values of $n\sigma^{\rm TOF}$, that reflects the TOF detector time response~\cite{Akindinov:2013tea}.

Particularly for this analysis, a special reconstruction of the Pb--Pb collisions was used, in order to minimise the sensitivity to elastic scattering in the material. In the TOF, tracks are only associated with a hit cluster if the cluster is within a certain distance from the extrapolated position of the track reconstructed in the TPC. This maximum distance was increased in this reconstruction, from 3 cm to 10 cm, with the latter being the usual window for pp collisions.

\subsection{Corrections}
\label{Sect:corrections}

The sample of $^{3}$H candidates reconstructed as described above includes both the nuclei produced in the initial collisions and those produced in the spallation processes occurring when particles traverse the beam pipe and the detector material. Since this process produces nuclei by spallation from other heavier nuclei, it can not produce antinuclei, but rather secondary nuclei only. The antibaryon-to-baryon ratio method employed in pp collisions is sensitive to these secondary nuclei, since it compares the yields of antitriton to those of triton in order to measure the inelastic cross section. Therefore, this effect needs to be corrected for. In order to distinguish between tritons originating from the initial collision and those originating from spallation processes, their different DCAs to the primary vertex in the transverse plane are used. While primary nuclei have DCA distributions peaked around zero, secondary nuclei show much broader DCA distributions which are flat over most of the studied DCA range. The different DCA distributions for the two types of nuclei are studied with templates from detailed MC simulations, and the relative contributions are therefore obtained, as shown in details in Ref.~\cite{3He_pPb}. The uncertainty from this correction is estimated to be below 5\%. 

A second correction is applied to account for any energy loss which occurs before the inelastic interactions, due to ionisation and rescattering processes. The energy loss has to be estimated differently between the two approaches, as what needs to be corrected in the antibaryon-to-baryon method is the energy loss between the primary vertex and the point where the \AntiHThree inelastic interaction occurs, while for the TOF-to-TPC method it is the energy loss between the TPC and the interaction point. For the antibaryon-to-baryon method, the momentum of the tracks is estimated by measuring their curvature, and it is then propagated towards the primary vertex accounting for the measured energy loss, and thus is the momentum at the primary vertex \pvert.

Due to continuous energy-loss effects in the detector material, the inelastic interaction of the antinuclei with the detector material happens at a momentum $p$, which is lower than the momentum \pvert reconstructed at the primary collision vertex. The corresponding effect is taken into account utilising MC simulations in which one has precise information about both momenta for each (anti)particle. In the analysis of pp collisions using the antibaryon-to-baryon method, the average values of $p/$\pvert distributions in each analysed \pvert interval are used to consider the energy loss. The root mean square (RMS) of these distributions is used to determine the uncertainty of the momentum p, which is propagated to the uncertainty of the measured cross section. For the analysis of the Pb–Pb data sample using the TOF-to-TPC method, the MC information on the momenta of daughter tracks originating from the \AntiHThree annihilation is used to estimate the momentum of the particle at annihilation. This is compared to \pvert to estimate the magnitude of this effect and the resulting uncertainty. The uncertainty on the inelastic cross section from the continous energy loss is evaluated to be less than 2\% for the antibaryon-to-baryon method and less than 2.5\% for the TOF-to-TPC method.

\subsection{Systematic uncertainties}
\label{Sect:syst}
Several sources of systematic uncertainties have been considered, depending on the method used for the inelastic cross section extraction. The total uncertainty is obtained as the quadratic sum of the individual contributions, assuming uncorrelated contributions. 

The first source of systematic uncertainty, investigated for both methods, is due to the track selection criteria. The criteria have been varied 100 times, both in data and MC, using random uniform distributions around the nominal values. In the case of the antibaryon-to-baryon method, the relative systematic uncertainty is given by the RMS divided by the mean value of the distributions of the \AntiHThree-to-$^3$H ratios in each momentum interval. For the TOF-to-TPC ratio method, the systematic uncertainty is evaluated as half the maximum difference between the resulting inelastic cross sections.
Variations consistent with statistical fluctuations are rejected from the trials used to estimate the systematic uncertainties according to the Barlow criterion~\cite{Barlow:2002yb}. This contribution is 13$\%$ for the TOF-to-TPC method, flat in momentum, while it is rejected by the Barlow test in the other method, using a cutoff value of $2\sigma_{\mathrm{Barlow}}$. This difference in the systematic uncertainty due to track selection criteria is due to the fact that in the antibaryon-to-baryon method, the effect of the track selection applies similarly to both the baryon and antibaryon tracks, and thus cancels out to a large extent. 

For the antibaryon-to-baryon ratio method, the systematic uncertainty due to the effect of secondary nuclei from spallation on the ratio is considered. This correction is described in section~\ref{Sect:corrections}. Variations of the binning and of the fit range (from $\pm1$ cm to $\pm2$ cm around the primary vertex) have been performed to evaluate this uncertainty, which is $\sim$6\% in the lowest momentum bin and $<1$\% at higher momenta. For the antibaryon-to-baryon method~\cite{antideuteronXS}, three additional uncertainties are included: i) the uncertainty on the primordial antimatter-to-matter ratio produced in collisions, ii) the uncertainty due to the elastic cross section of $^3$H and iii) the one due to the inelastic cross section of $^3$H. All three are considered as global uncertainties for the antibaryon-to-baryon method. These uncertainties are 4.5\%, 1\%, and 2.3\% in the whole momentum interval, respectively.

For the TOF-to-TPC ratio method, additional sources of systematic uncertainties are related to the description of the material budget in the MC simulations, the PID, and the momentum correction evaluation. 
The material budget of ALICE is varied by $\pm 4.5\%$ in MC simulations, and the deviations of the final results from the default case are considered as an uncertainty. The value of 4.5\% corresponds to the current uncertainty on the material obtained by photon conversion measurements~\cite{Abelev:2014ffa} and by the studies of the TOF-to-TPC matching efficiency with pions~\cite{ALICE-PUBLIC-2022-001}. The uncertainty related to the PID has been evaluated as the difference between the raw yields obtained by subtracting the background contribution from the $n\sigma^{\rm TPC}$ distributions when varying the function used to fit the contamination (Gaussian, double Gaussian, Gaussian + 2 exponential functions). The contribution coming from the difference between the bin-counting method and the integral of the signal function is added in quadrature to the previous one. The deviation of the final results from the default case divided by $\sqrt{12}$ is considered as uncertainty. 

The systematic uncertainty due to the energy loss correction is evaluated as described in section~\ref{Sect:corrections}, and amounts to $\sim$2\% ($\sim5$\%) in the analysis of Pb--Pb (pp) collisions.

\subsection{Monte Carlo simulations}
\label{Sect:MC}
\begin{figure}[!hbt]
\centering
\includegraphics[width=0.49\linewidth]{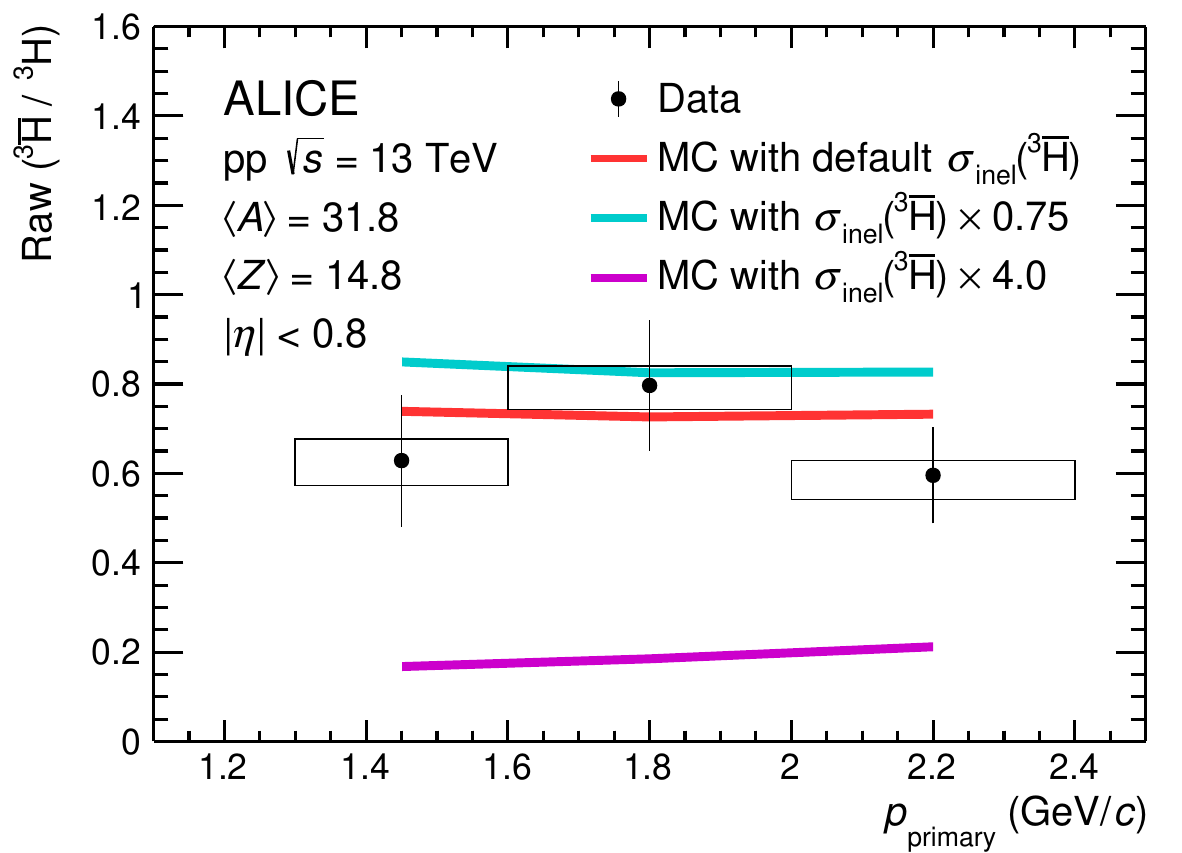}
\includegraphics[width=0.49\linewidth]{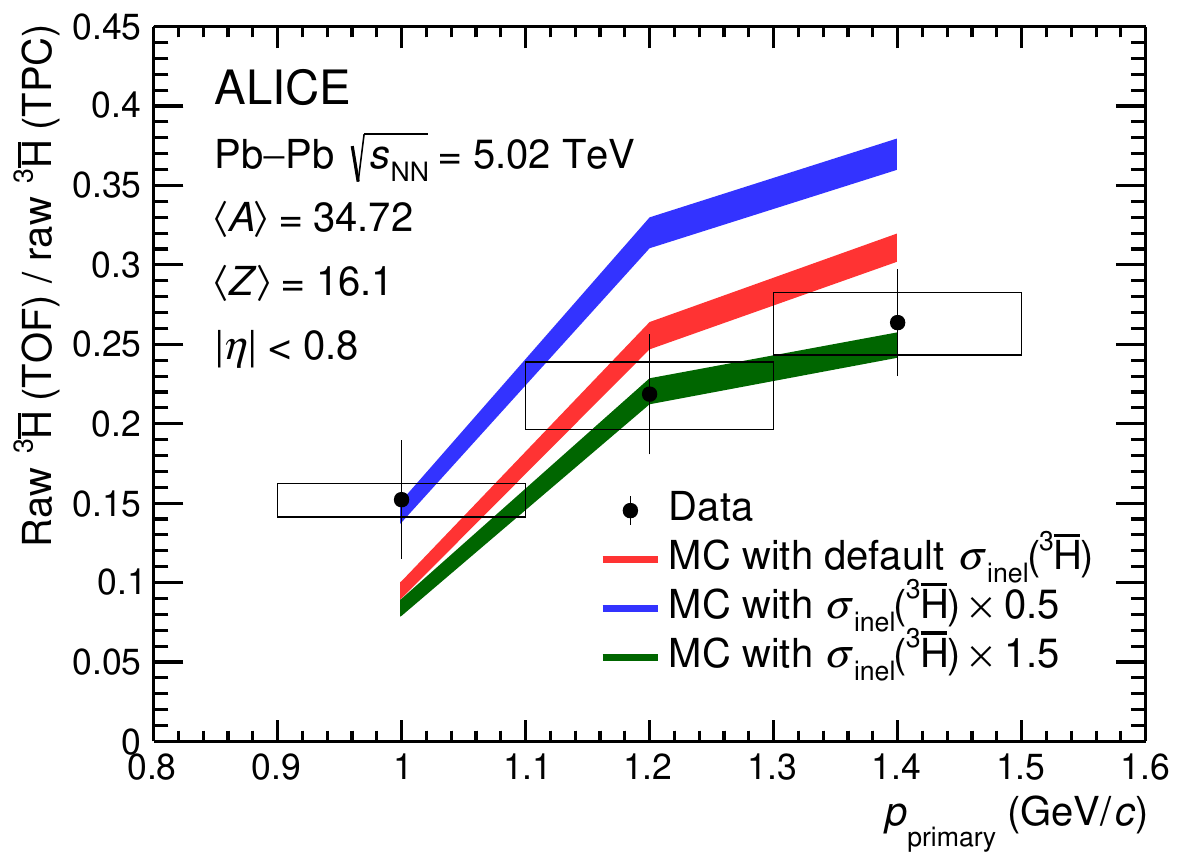}
\hfill
\caption{Left: raw primary \RatioAntit ratio as a function of the momentum $\pvert$ in pp collisions at $\sqrt{s} = 13$ TeV. Right: ratio between the raw number of \AntiHThree candidates reconstructed in the TOF and the raw number of \AntiHThree reconstructed in the TPC in Pb--Pb collisions at \snn = 5.02 TeV as a function of the momentum $\pvert$. In both panels, data are shown in black, the statistical and systematic uncertainties are shown as vertical bars and boxes, respectively. The results from ALICE MC simulations based on \geantf are shown as coloured bands, the different colours referring to the different inelastic cross sections implemented in the simulations, with default $\sigma_{\rm inel}$ described in Ref.~\cite{Uzhinsky:2011zz}. The width of the MC band represents the statistical uncertainty of the simulation.}
\label{fig:RawRatios}
\end{figure}

The experimental observables, shown in Fig.~\ref{fig:RawRatios}, are compared with the corresponding ratios from MC simulations based on \geantf. The parameterisation of the inelastic cross section in \geantf is based on Glauber calculations~\cite{GRICHINE201860, Uzhinsky:2011zz}, which are based on geometrically scaling the inelastic antiproton--proton cross sections to heavier systems. This parameterisation neglects the effect of Coulomb interactions, which are negligible for $p > 1$ \GeVc.
The analysis described above is repeated using several MC simulations with varied values of the antitriton inelastic cross sections. In the case of the antibaryon-to-baryon method, three MC simulations are used, with \SigmaInelAntiT multiplied by a factor of 0.75, 1, and 4. The simulations with these factors were already available and were thus used. In the case of the TOF-to-TPC method, three MC simulations are used, with \SigmaInelAntiT multiplied by a factor of 0.5 and 1.5, in addition to the default one.
The corresponding MC ratios for the two cases are shown as coloured bands in Fig.~\ref{fig:RawRatios} and are used as references for the experimental ratios to obtain the \AntiHThree inelastic cross section as described in the following section.

\subsection{Inelastic cross section determination}
\label{Sect:inelXS}
\begin{figure}[hbt]
\begin{minipage}[h]{1\linewidth}
\center{\includegraphics[width=.49\linewidth]{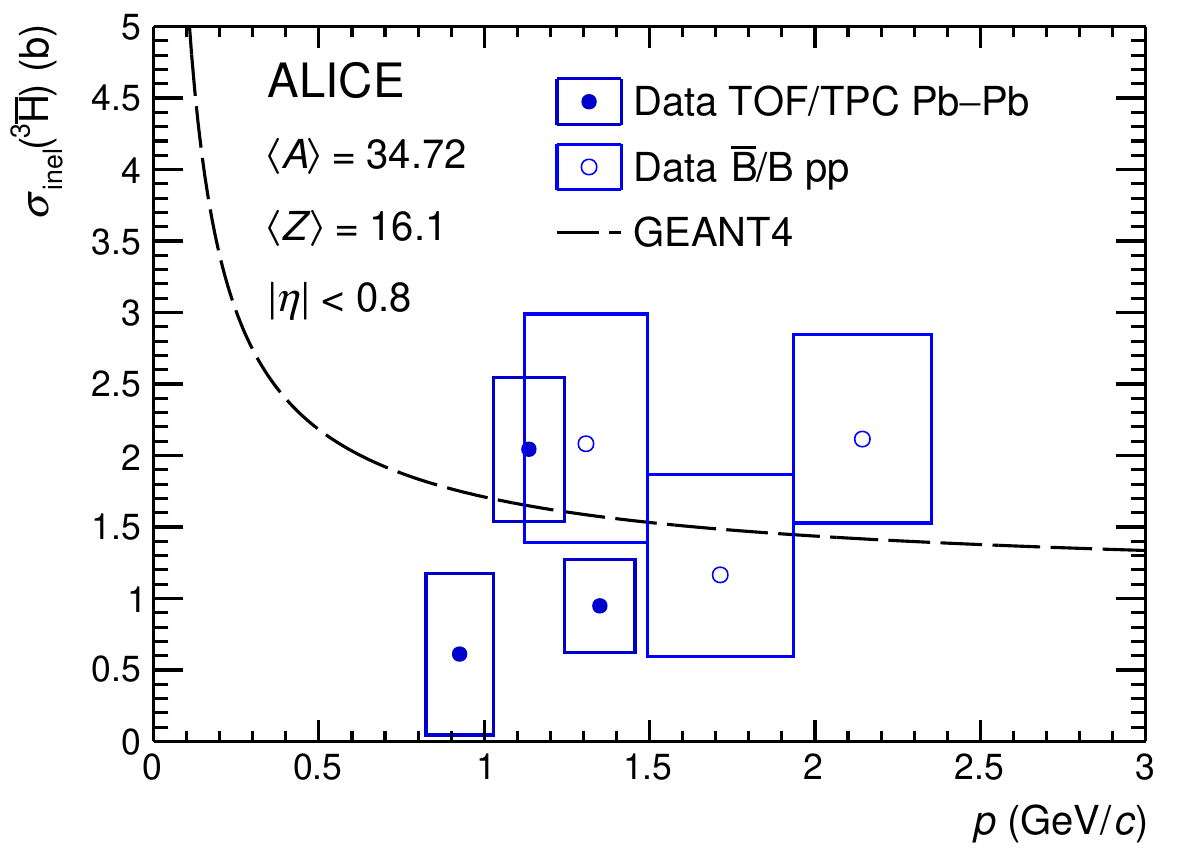} \includegraphics[width=.49\linewidth]{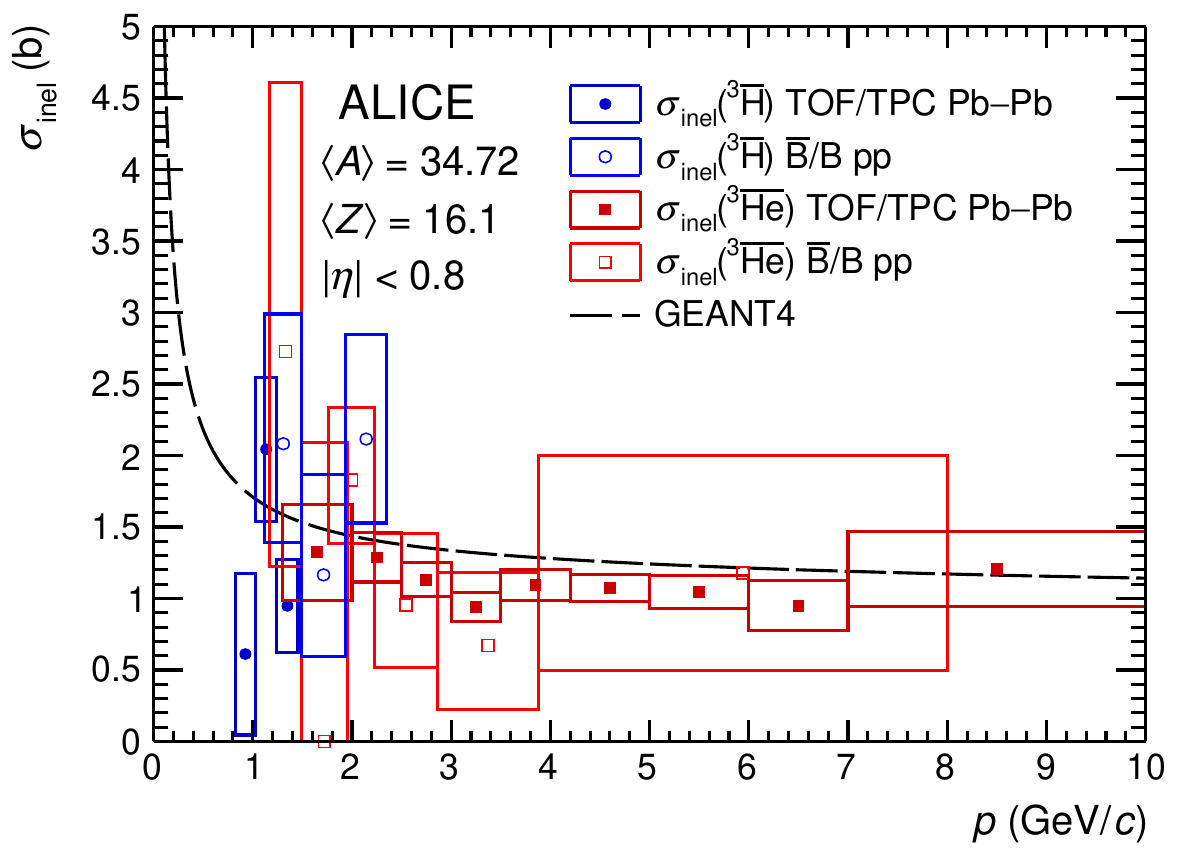}}
\end{minipage}
\caption{Left: inelastic interaction cross section for antitritons with the two analysis methods on an average material element of the ALICE detector as a function of the momentum $p$ at which the interaction occurs. Dashed black lines represent the default \geantf parameterisations for antitritons. Right: comparison between the results reported in the left panel and the results for \AntiHeThree inelastic cross section measured in Pb--Pb collisions at \snn = 5.02 TeV from Ref.~\cite{antiHe3XS}. Note that \AntiHeThree nuclei can be clearly identified in both TPC and TOF detectors over a much wider momentum range than \antit.
In both panels, boxes show the statistical and systematic uncertainties summed in quadrature.}
\label{fig:CrossSections_all}
\end{figure}

The determination of the inelastic cross section requires precise knowledge of the ALICE detector material budget. Detailed studies of the detector material have been carried out in previous works by the ALICE Collaboration~\cite{antideuteronXS, antiHe3XS}. These studies allow the determination of the effective target material. 
For the analysis based on the antibaryon-to-baryon method, the inelastic interactions can occur in the ITS, TPC, TRD, and TOF detectors, hence the average material corresponds to $\langle Z\rangle = 14.8$ and $\langle A\rangle = 31.8$. For the TOF-to-TPC method, instead, the average material is the TRD and space frame system, corresponding to $\langle Z\rangle = 16.1$ and $\langle A\rangle = 34.7$. 
Using the same procedure described in Refs.~\cite{antideuteronXS, antiHe3XS}, the MC ratios obtained with the different multiplicative factors applied to $\sigma_{\mathrm{inel}}($\AntiHThree) are fitted in each momentum interval using a Lambert--Beer function: 
\begin{equation}
    R = \exp(- \rm a \sigma ) \times b,
\end{equation} 
where $R$ refers to the MC ratios obtained with either of the two methods, a and b are the fit parameters, and $\sigma$ refers to the scaling factor of the \SigmaInelAntiT. 
The experimental ratios are therefore projected on the fit function and a scaling factor for the \geantf parameterisation is obtained for each momentum interval.  
This scaling factor shows by how much the inelastic cross section needs to be changed in the \geantf MC implementation with respect to default to reach the same value for the observable (antitriton-to-triton ratio for pp, TPC-to-TOF ratio for Pb--Pb) as in the data. Finally, by multiplying the scaling factor in each interval by the default inelastic cross section implemented in \geantf, the inelastic cross section is obtained. Since only cross sections on integer atomic numbers can be obtained in \geantf, the inelastic cross section on the nearest available element was scaled according to the Glauber model~\cite{GRICHINE201860} as described in equation~\ref{eq:AScaling} 
\begin{equation}\label{eq:AScaling}
  \sigma(\rm{A_{ALICE}})= \sigma(\rm{A_{\textsc{Geant4}}}) \times 
  \frac{1.34 \times \rm{A_{\textsc{Geant4}}^{0.21}} + 1.51 \times \rm{A_{\textsc{Geant4}}^{-1/3}}} {1.34 \times \rm{A_{ALICE}^{0.21}} + 1.51 \times \rm{A_{ALICE}^{-1/3}}}
\end{equation}
where $\rm{A_{ALICE}}$ is the average atomic number seen by antinuclei as they travel through the ALICE detector and $\rm{A_{\textsc{Geant4}}}$ is the atomic number of the closest available element in \geantf.
The resulting \SigmaInelAntiT is shown in the left panel of Fig.~\ref{fig:CrossSections_all}, for the two analysed data samples. For the ease of comparison, the results from the antibaryon-to-baryon analysis have been scaled to the same average material as the results from the TOF-to-TPC analysis, also according to equation~\ref{eq:AScaling}, and $\sigma($\geantf) is shown as well. Since $\sigma($\geantf) does not depend on the collision system, this allows for a fair comparison of the shown measurements. It can be seen that the measurements using the antibaryon-to-baryon method agree with $\sigma($\geantf) at a 1 $\sigma$ level, and the TOF-to-TPC ratio method at a 2 $\sigma$ level.
The inelastic cross sections are shown as a function of the momentum $p$ at which the inelastic interaction occurs.\\ 
Comparing the inelastic cross sections for $^3\overline{\rm H}$ and $^3\overline{\rm{He}}$ shows that the two measurements are consistent within uncertainties (see right panel of Fig.~\ref{fig:CrossSections_all}). The uncertainties are dominated by statistical uncertainties, and thus do not cancel when taking the ratio of \SigmaInelAntiT and $\sigma_{\mathrm{inel}}(^3\overline{\mathrm{He}})$. There are no quantitative predictions for the expected difference between the two isospin partners, however, a naive approach would be to consider their size as the dominant factor. No measurements of the nuclear mass radii of these antinuclei exist, and thus the matter radii are taken as a proxy. However, since the difference in mass radii is expected to be smaller than the difference in matter radii, this is a conservative estimate. The measured matter radii of the two nuclei are 1.76~fm for $^3$He \cite{Ropke:2008qk} and 1.59~fm for $^3$H \cite{Ropke:2008qk}. While no independent measurement exists for the antinuclei counterparts, the antinuclei are assumed to have the same size as their matter counterparts by CPT symmetry. This would suggest that the inelastic cross section of \ahe\ should be a factor 0 -- 20\% larger than the one of \antit. The current measurement is not sensitive enough to distinguish differences of that order and the cross sections of \AntiHThree and \AntiHeThree are compatible within errors. The larger size of the data samples expected in the upcoming Run 3 and Run 4 campaigns of the LHC will allow a significant improvement over the current measurement.

\section{Summary}
\label{Sect:Results}

The results presented in this Letter represent the first measurement of the antitriton inelastic cross section, and complete the set of measurements of antinuclei--matter inelastic cross section up to $A=3$ of the projectile antinucleus by ALICE. The results have been found, within sizeable uncertainties, to be consistent with the parameterisation used in \geantf toolkit and with existing \ahe\ measurements from Ref.~\cite{antiHe3XS}. They demonstrate the feasibility of studying the isospin dependence of inelastic interactions of composite antinuclei such as \antit and \ahe. Future studies of larger data samples collected during the LHC Run 3 and Run 4 campaigns (starting from 2022) with the upgraded ALICE apparatus will substantially improve on the measurements presented in this Letter. They will allow for more precise comparison between \antit\ and \ahe\ results and for better calculations of the inelastic cross sections of antinuclei in the currently used models, improving also our understanding of antinucleus--matter inelastic interactions at momenta around 1 GeV/$c$.

\newenvironment{acknowledgement}{\relax}{\relax}
\begin{acknowledgement}
\section*{Acknowledgements}

The ALICE Collaboration would like to thank all its engineers and technicians for their invaluable contributions to the construction of the experiment and the CERN accelerator teams for the outstanding performance of the LHC complex.
The ALICE Collaboration gratefully acknowledges the resources and support provided by all Grid centres and the Worldwide LHC Computing Grid (WLCG) collaboration.
The ALICE Collaboration acknowledges the following funding agencies for their support in building and running the ALICE detector:
A. I. Alikhanyan National Science Laboratory (Yerevan Physics Institute) Foundation (ANSL), State Committee of Science and World Federation of Scientists (WFS), Armenia;
Austrian Academy of Sciences, Austrian Science Fund (FWF): [M 2467-N36] and Nationalstiftung f\"{u}r Forschung, Technologie und Entwicklung, Austria;
Ministry of Communications and High Technologies, National Nuclear Research Center, Azerbaijan;
Conselho Nacional de Desenvolvimento Cient\'{\i}fico e Tecnol\'{o}gico (CNPq), Financiadora de Estudos e Projetos (Finep), Funda\c{c}\~{a}o de Amparo \`{a} Pesquisa do Estado de S\~{a}o Paulo (FAPESP) and Universidade Federal do Rio Grande do Sul (UFRGS), Brazil;
Bulgarian Ministry of Education and Science, within the National Roadmap for Research Infrastructures 2020-2027 (object CERN), Bulgaria;
Ministry of Education of China (MOEC) , Ministry of Science \& Technology of China (MSTC) and National Natural Science Foundation of China (NSFC), China;
Ministry of Science and Education and Croatian Science Foundation, Croatia;
Centro de Aplicaciones Tecnol\'{o}gicas y Desarrollo Nuclear (CEADEN), Cubaenerg\'{\i}a, Cuba;
Ministry of Education, Youth and Sports of the Czech Republic, Czech Republic;
The Danish Council for Independent Research | Natural Sciences, the VILLUM FONDEN and Danish National Research Foundation (DNRF), Denmark;
Helsinki Institute of Physics (HIP), Finland;
Commissariat \`{a} l'Energie Atomique (CEA) and Institut National de Physique Nucl\'{e}aire et de Physique des Particules (IN2P3) and Centre National de la Recherche Scientifique (CNRS), France;
Bundesministerium f\"{u}r Bildung und Forschung (BMBF) and GSI Helmholtzzentrum f\"{u}r Schwerionenforschung GmbH, Germany;
General Secretariat for Research and Technology, Ministry of Education, Research and Religions, Greece;
National Research, Development and Innovation Office, Hungary;
Department of Atomic Energy Government of India (DAE), Department of Science and Technology, Government of India (DST), University Grants Commission, Government of India (UGC) and Council of Scientific and Industrial Research (CSIR), India;
National Research and Innovation Agency - BRIN, Indonesia;
Istituto Nazionale di Fisica Nucleare (INFN), Italy;
Japanese Ministry of Education, Culture, Sports, Science and Technology (MEXT) and Japan Society for the Promotion of Science (JSPS) KAKENHI, Japan;
Consejo Nacional de Ciencia (CONACYT) y Tecnolog\'{i}a, through Fondo de Cooperaci\'{o}n Internacional en Ciencia y Tecnolog\'{i}a (FONCICYT) and Direcci\'{o}n General de Asuntos del Personal Academico (DGAPA), Mexico;
Nederlandse Organisatie voor Wetenschappelijk Onderzoek (NWO), Netherlands;
The Research Council of Norway, Norway;
Commission on Science and Technology for Sustainable Development in the South (COMSATS), Pakistan;
Pontificia Universidad Cat\'{o}lica del Per\'{u}, Peru;
Ministry of Education and Science, National Science Centre and WUT ID-UB, Poland;
Korea Institute of Science and Technology Information and National Research Foundation of Korea (NRF), Republic of Korea;
Ministry of Education and Scientific Research, Institute of Atomic Physics, Ministry of Research and Innovation and Institute of Atomic Physics and University Politehnica of Bucharest, Romania;
Ministry of Education, Science, Research and Sport of the Slovak Republic, Slovakia;
National Research Foundation of South Africa, South Africa;
Swedish Research Council (VR) and Knut \& Alice Wallenberg Foundation (KAW), Sweden;
European Organization for Nuclear Research, Switzerland;
Suranaree University of Technology (SUT), National Science and Technology Development Agency (NSTDA), Thailand Science Research and Innovation (TSRI) and National Science, Research and Innovation Fund (NSRF), Thailand;
Turkish Energy, Nuclear and Mineral Research Agency (TENMAK), Turkey;
National Academy of  Sciences of Ukraine, Ukraine;
Science and Technology Facilities Council (STFC), United Kingdom;
National Science Foundation of the United States of America (NSF) and United States Department of Energy, Office of Nuclear Physics (DOE NP), United States of America.
In addition, individual groups or members have received support from:
European Research Council, Strong 2020 - Horizon 2020 (grant nos. 950692, 824093), European Union;
Academy of Finland (Center of Excellence in Quark Matter) (grant nos. 346327, 346328), Finland.
\end{acknowledgement}

\bibliographystyle{utphys}   
\bibliography{bibliography}

\newpage
\appendix

\newpage

\section{The ALICE Collaboration}
\label{app:collab}
\begin{flushleft} 
\small

S.~Acharya\,\orcidlink{0000-0002-9213-5329}\,$^{\rm 127}$, 
D.~Adamov\'{a}\,\orcidlink{0000-0002-0504-7428}\,$^{\rm 87}$, 
A.~Adler$^{\rm 71}$, 
G.~Aglieri Rinella\,\orcidlink{0000-0002-9611-3696}\,$^{\rm 33}$, 
M.~Agnello\,\orcidlink{0000-0002-0760-5075}\,$^{\rm 30}$, 
N.~Agrawal\,\orcidlink{0000-0003-0348-9836}\,$^{\rm 52}$, 
Z.~Ahammed\,\orcidlink{0000-0001-5241-7412}\,$^{\rm 135}$, 
S.~Ahmad\,\orcidlink{0000-0003-0497-5705}\,$^{\rm 16}$, 
S.U.~Ahn\,\orcidlink{0000-0001-8847-489X}\,$^{\rm 72}$, 
I.~Ahuja\,\orcidlink{0000-0002-4417-1392}\,$^{\rm 38}$
% ...

\section*{Affiliation Notes}

$^{\rm I}$ Deceased\\
$^{\rm II}$ Also at: Max-Planck-Institut fur Physik, Munich, Germany\\
$^{\rm III}$ Also at: Italian National Agency for New Technologies, Energy and Sustainable Economic Development (ENEA), Bologna, Italy\\
$^{\rm IV}$ Also at: Dipartimento DET del Politecnico di Torino, Turin, Italy\\
$^{\rm V}$ Also at: Department of Applied Physics, Aligarh Muslim University, Aligarh, India\\
$^{\rm VI}$ Also at: Institute of Theoretical Physics, University of Wroclaw, Poland\\
$^{\rm VII}$ Also at: An institution covered by a cooperation agreement with CERN\\

\section*{Collaboration Institutes}

$^{1}$ A.I. Alikhanyan National Science Laboratory (Yerevan Physics Institute) Foundation, Yerevan, Armenia\\
$^{2}$ AGH University of Krakow, Cracow, Poland\\
$^{3}$ Bogolyubov Institute for Theoretical Physics, National Academy of Sciences of Ukraine, Kiev, Ukraine\\
$^{4}$ Bose Institute, Department of Physics  and Centre for Astroparticle Physics and Space Science (CAPSS), Kolkata, India\\
$^{5}$ California Polytechnic State University, San Luis Obispo, California, United States\\
$^{6}$ Central China Normal University, Wuhan, China\\
$^{7}$ Centro de Aplicaciones Tecnol\'{o}gicas y Desarrollo Nuclear (CEADEN), Havana, Cuba\\
$^{8}$ Centro de Investigaci\'{o}n y de Estudios Avanzados (CINVESTAV), Mexico City and M\'{e}rida, Mexico\\
$^{9}$ Chicago State University, Chicago, Illinois, United States\\
$^{10}$ China Institute of Atomic Energy, Beijing, China\\
$^{11}$ China University of Geosciences, Wuhan, China\\
$^{12}$ Chungbuk National University, Cheongju, Republic of Korea\\
$^{13}$ Comenius University Bratislava, Faculty of Mathematics, Physics and Informatics, Bratislava, Slovak Republic\\
$^{14}$ COMSATS University Islamabad, Islamabad, Pakistan\\
$^{15}$ Creighton University, Omaha, Nebraska, United States\\
$^{16}$ Department of Physics, Aligarh Muslim University, Aligarh, India\\
$^{17}$ Department of Physics, Pusan National University, Pusan, Republic of Korea\\
$^{18}$ Department of Physics, Sejong University, Seoul, Republic of Korea\\
$^{19}$ Department of Physics, University of California, Berkeley, California, United States\\
$^{20}$ Department of Physics, University of Oslo, Oslo, Norway\\
$^{21}$ Department of Physics and Technology, University of Bergen, Bergen, Norway\\
$^{22}$ Dipartimento di Fisica, Universit\`{a} di Pavia, Pavia, Italy\\
$^{23}$ Dipartimento di Fisica dell'Universit\`{a} and Sezione INFN, Cagliari, Italy\\
$^{24}$ Dipartimento di Fisica dell'Universit\`{a} and Sezione INFN, Trieste, Italy\\
$^{25}$ Dipartimento di Fisica dell'Universit\`{a} and Sezione INFN, Turin, Italy\\
$^{26}$ Dipartimento di Fisica e Astronomia dell'Universit\`{a} and Sezione INFN, Bologna, Italy\\
$^{27}$ Dipartimento di Fisica e Astronomia dell'Universit\`{a} and Sezione INFN, Catania, Italy\\
$^{28}$ Dipartimento di Fisica e Astronomia dell'Universit\`{a} and Sezione INFN, Padova, Italy\\
$^{29}$ Dipartimento di Fisica `E.R.~Caianiello' dell'Universit\`{a} and Gruppo Collegato INFN, Salerno, Italy\\
$^{30}$ Dipartimento DISAT del Politecnico and Sezione INFN, Turin, Italy\\
$^{31}$ Dipartimento di Scienze MIFT, Universit\`{a} di Messina, Messina, Italy\\
$^{32}$ Dipartimento Interateneo di Fisica `M.~Merlin' and Sezione INFN, Bari, Italy\\
$^{33}$ European Organization for Nuclear Research (CERN), Geneva, Switzerland\\
$^{34}$ Faculty of Electrical Engineering, Mechanical Engineering and Naval Architecture, University of Split, Split, Croatia\\
$^{35}$ Faculty of Engineering and Science, Western Norway University of Applied Sciences, Bergen, Norway\\
$^{36}$ Faculty of Nuclear Sciences and Physical Engineering, Czech Technical University in Prague, Prague, Czech Republic\\
$^{37}$ Faculty of Physics, Sofia University, Sofia, Bulgaria\\
$^{38}$ Faculty of Science, P.J.~\v{S}af\'{a}rik University, Ko\v{s}ice, Slovak Republic\\
$^{39}$ Frankfurt Institute for Advanced Studies, Johann Wolfgang Goethe-Universit\"{a}t Frankfurt, Frankfurt, Germany\\
$^{40}$ Fudan University, Shanghai, China\\
$^{41}$ Gangneung-Wonju National University, Gangneung, Republic of Korea\\
$^{42}$ Gauhati University, Department of Physics, Guwahati, India\\
$^{43}$ Helmholtz-Institut f\"{u}r Strahlen- und Kernphysik, Rheinische Friedrich-Wilhelms-Universit\"{a}t Bonn, Bonn, Germany\\
$^{44}$ Helsinki Institute of Physics (HIP), Helsinki, Finland\\
$^{45}$ High Energy Physics Group,  Universidad Aut\'{o}noma de Puebla, Puebla, Mexico\\
$^{46}$ Horia Hulubei National Institute of Physics and Nuclear Engineering, Bucharest, Romania\\
$^{47}$ HUN-REN Wigner Research Centre for Physics, Budapest, Hungary\\
$^{48}$ Indian Institute of Technology Bombay (IIT), Mumbai, India\\
$^{49}$ Indian Institute of Technology Indore, Indore, India\\
$^{50}$ INFN, Laboratori Nazionali di Frascati, Frascati, Italy\\
$^{51}$ INFN, Sezione di Bari, Bari, Italy\\
$^{52}$ INFN, Sezione di Bologna, Bologna, Italy\\
$^{53}$ INFN, Sezione di Cagliari, Cagliari, Italy\\
$^{54}$ INFN, Sezione di Catania, Catania, Italy\\
$^{55}$ INFN, Sezione di Padova, Padova, Italy\\
$^{56}$ INFN, Sezione di Pavia, Pavia, Italy\\
$^{57}$ INFN, Sezione di Torino, Turin, Italy\\
$^{58}$ INFN, Sezione di Trieste, Trieste, Italy\\
$^{59}$ Inha University, Incheon, Republic of Korea\\
$^{60}$ Institute for Gravitational and Subatomic Physics (GRASP), Utrecht University/Nikhef, Utrecht, Netherlands\\
$^{61}$ Institute of Experimental Physics, Slovak Academy of Sciences, Ko\v{s}ice, Slovak Republic\\
$^{62}$ Institute of Physics, Homi Bhabha National Institute, Bhubaneswar, India\\
$^{63}$ Institute of Physics of the Czech Academy of Sciences, Prague, Czech Republic\\
$^{64}$ Institute of Space Science (ISS), Bucharest, Romania\\
$^{65}$ Institut f\"{u}r Kernphysik, Johann Wolfgang Goethe-Universit\"{a}t Frankfurt, Frankfurt, Germany\\
$^{66}$ Instituto de Ciencias Nucleares, Universidad Nacional Aut\'{o}noma de M\'{e}xico, Mexico City, Mexico\\
$^{67}$ Instituto de F\'{i}sica, Universidade Federal do Rio Grande do Sul (UFRGS), Porto Alegre, Brazil\\
$^{68}$ Instituto de F\'{\i}sica, Universidad Nacional Aut\'{o}noma de M\'{e}xico, Mexico City, Mexico\\
$^{69}$ iThemba LABS, National Research Foundation, Somerset West, South Africa\\
$^{70}$ Jeonbuk National University, Jeonju, Republic of Korea\\
$^{71}$ Johann-Wolfgang-Goethe Universit\"{a}t Frankfurt Institut f\"{u}r Informatik, Fachbereich Informatik und Mathematik, Frankfurt, Germany\\
$^{72}$ Korea Institute of Science and Technology Information, Daejeon, Republic of Korea\\
$^{73}$ KTO Karatay University, Konya, Turkey\\
$^{74}$ Laboratoire de Physique Subatomique et de Cosmologie, Universit\'{e} Grenoble-Alpes, CNRS-IN2P3, Grenoble, France\\
$^{75}$ Lawrence Berkeley National Laboratory, Berkeley, California, United States\\
$^{76}$ Lund University Department of Physics, Division of Particle Physics, Lund, Sweden\\
$^{77}$ Nagasaki Institute of Applied Science, Nagasaki, Japan\\
$^{78}$ Nara Women{'}s University (NWU), Nara, Japan\\
$^{79}$ National and Kapodistrian University of Athens, School of Science, Department of Physics , Athens, Greece\\
$^{80}$ National Centre for Nuclear Research, Warsaw, Poland\\
$^{81}$ National Institute of Science Education and Research, Homi Bhabha National Institute, Jatni, India\\
$^{82}$ National Nuclear Research Center, Baku, Azerbaijan\\
$^{83}$ National Research and Innovation Agency - BRIN, Jakarta, Indonesia\\
$^{84}$ Niels Bohr Institute, University of Copenhagen, Copenhagen, Denmark\\
$^{85}$ Nikhef, National institute for subatomic physics, Amsterdam, Netherlands\\
$^{86}$ Nuclear Physics Group, STFC Daresbury Laboratory, Daresbury, United Kingdom\\
$^{87}$ Nuclear Physics Institute of the Czech Academy of Sciences, Husinec-\v{R}e\v{z}, Czech Republic\\
$^{88}$ Oak Ridge National Laboratory, Oak Ridge, Tennessee, United States\\
$^{89}$ Ohio State University, Columbus, Ohio, United States\\
$^{90}$ Physics department, Faculty of science, University of Zagreb, Zagreb, Croatia\\
$^{91}$ Physics Department, Panjab University, Chandigarh, India\\
$^{92}$ Physics Department, University of Jammu, Jammu, India\\
$^{93}$ Physics Program and International Institute for Sustainability with Knotted Chiral Meta Matter (SKCM2), Hiroshima University, Hiroshima, Japan\\
$^{94}$ Physikalisches Institut, Eberhard-Karls-Universit\"{a}t T\"{u}bingen, T\"{u}bingen, Germany\\
$^{95}$ Physikalisches Institut, Ruprecht-Karls-Universit\"{a}t Heidelberg, Heidelberg, Germany\\
$^{96}$ Physik Department, Technische Universit\"{a}t M\"{u}nchen, Munich, Germany\\
$^{97}$ Politecnico di Bari and Sezione INFN, Bari, Italy\\
$^{98}$ Research Division and ExtreMe Matter Institute EMMI, GSI Helmholtzzentrum f\"ur Schwerionenforschung GmbH, Darmstadt, Germany\\
$^{99}$ Saga University, Saga, Japan\\
$^{100}$ Saha Institute of Nuclear Physics, Homi Bhabha National Institute, Kolkata, India\\
$^{101}$ School of Physics and Astronomy, University of Birmingham, Birmingham, United Kingdom\\
$^{102}$ Secci\'{o}n F\'{\i}sica, Departamento de Ciencias, Pontificia Universidad Cat\'{o}lica del Per\'{u}, Lima, Peru\\
$^{103}$ Stefan Meyer Institut f\"{u}r Subatomare Physik (SMI), Vienna, Austria\\
$^{104}$ SUBATECH, IMT Atlantique, Nantes Universit\'{e}, CNRS-IN2P3, Nantes, France\\
$^{105}$ Sungkyunkwan University, Suwon City, Republic of Korea\\
$^{106}$ Suranaree University of Technology, Nakhon Ratchasima, Thailand\\
$^{107}$ Technical University of Ko\v{s}ice, Ko\v{s}ice, Slovak Republic\\
$^{108}$ The Henryk Niewodniczanski Institute of Nuclear Physics, Polish Academy of Sciences, Cracow, Poland\\
$^{109}$ The University of Texas at Austin, Austin, Texas, United States\\
$^{110}$ Universidad Aut\'{o}noma de Sinaloa, Culiac\'{a}n, Mexico\\
$^{111}$ Universidade de S\~{a}o Paulo (USP), S\~{a}o Paulo, Brazil\\
$^{112}$ Universidade Estadual de Campinas (UNICAMP), Campinas, Brazil\\
$^{113}$ Universidade Federal do ABC, Santo Andre, Brazil\\
$^{114}$ Universitatea Nationala de Stiinta si Tehnologie Politehnica Bucuresti, Bucharest, Romania\\
$^{115}$ University of Cape Town, Cape Town, South Africa\\
$^{116}$ University of Houston, Houston, Texas, United States\\
$^{117}$ University of Jyv\"{a}skyl\"{a}, Jyv\"{a}skyl\"{a}, Finland\\
$^{118}$ University of Kansas, Lawrence, Kansas, United States\\
$^{119}$ University of Liverpool, Liverpool, United Kingdom\\
$^{120}$ University of Science and Technology of China, Hefei, China\\
$^{121}$ University of South-Eastern Norway, Kongsberg, Norway\\
$^{122}$ University of Tennessee, Knoxville, Tennessee, United States\\
$^{123}$ University of the Witwatersrand, Johannesburg, South Africa\\
$^{124}$ University of Tokyo, Tokyo, Japan\\
$^{125}$ University of Tsukuba, Tsukuba, Japan\\
$^{126}$ Universit\"{a}t M\"{u}nster, Institut f\"{u}r Kernphysik, M\"{u}nster, Germany\\
$^{127}$ Universit\'{e} Clermont Auvergne, CNRS/IN2P3, LPC, Clermont-Ferrand, France\\
$^{128}$ Universit\'{e} de Lyon, CNRS/IN2P3, Institut de Physique des 2 Infinis de Lyon, Lyon, France\\
$^{129}$ Universit\'{e} de Strasbourg, CNRS, IPHC UMR 7178, F-67000 Strasbourg, France, Strasbourg, France\\
$^{130}$ Universit\'{e} Paris-Saclay, Centre d'Etudes de Saclay (CEA), IRFU, D\'{e}partment de Physique Nucl\'{e}aire (DPhN), Saclay, France\\
$^{131}$ Universit\'{e}  Paris-Saclay, CNRS/IN2P3, IJCLab, Orsay, France\\
$^{132}$ Universit\`{a} degli Studi di Foggia, Foggia, Italy\\
$^{133}$ Universit\`{a} del Piemonte Orientale, Vercelli, Italy\\
$^{134}$ Universit\`{a} di Brescia, Brescia, Italy\\
$^{135}$ Variable Energy Cyclotron Centre, Homi Bhabha National Institute, Kolkata, India\\
$^{136}$ Warsaw University of Technology, Warsaw, Poland\\
$^{137}$ Wayne State University, Detroit, Michigan, United States\\
$^{138}$ Yale University, New Haven, Connecticut, United States\\
$^{139}$ Yonsei University, Seoul, Republic of Korea\\
$^{140}$  Zentrum  f\"{u}r Technologie und Transfer (ZTT), Worms, Germany\\
$^{141}$ Affiliated with an institute covered by a cooperation agreement with CERN\\
$^{142}$ Affiliated with an international laboratory covered by a cooperation agreement with CERN.\\

\end{flushleft} 

\end{document}